\documentclass[twocolumn,prd,aps,showpacs,superscriptaddress]{revtex4}
\usepackage{amsmath}
\usepackage{graphicx}
\usepackage[usenames]{color}
\usepackage[letterpaper,margin=0.75in]{geometry}

\usepackage[colorlinks,bookmarks]{hyperref}
\definecolor{linkblue}{rgb}{0,0,0.8}
\definecolor{linkgreen}{rgb}{0,0.5,0}
\hypersetup{pdfpagemode=None, pdfstartview=FitH, linkcolor=linkblue, %
             citecolor=linkgreen, urlcolor=linkblue}

\def\beq{\begin{equation}}
\def\eeq{\end{equation}}
\def\bea{\setlength\arraycolsep{1.4pt}\begin{eqnarray}}
\def\eea{\end{eqnarray}}
\def\bit{\begin{itemize}}
\def\eit{\end{itemize}}
\def\nn{\nonumber}

\def\eq{Eq.~}
\def\eqs{Eqs.~}
\def\fig{Fig.~}

\def\pd{\partial}

\def\ld{\left}
\def\rd{\right}

\def\tl{\tilde}
\def\wtl{\widetilde}
\def\ph{\phantom}
\def\fr{\frac}
\def\oo{\frac{1}}
\def\half{\frac{1}{2}}

\def\del{\delta}
\def\Del{\Delta}

\def\Lam{\Lambda}

\def\sig{\sigma}

\def\Om{\Omega}

\def\L{{\cal L}}
\def\O{{\cal O}}

\def\R{{\cal R}}
\def\PR{{{\cal P}_{\cal R}}}
\def\PRh{{{\cal P}_{\cal R}^{\rm hi}}}
\def\PRl{{{\cal P}_{\cal R}^{\rm lo}}}
\def\PRL{{{\cal P}_{\cal R}^{\Lam\rm CDM}}}
\def\Rl{{{\cal R}^{\rm lo}}}
\def\Rh{{{\cal R}^{\rm hi}}}
\def\tlRl{{\wtl{\cal R}^{\rm lo}}}

\def\Sl{{S^{\rm lo}}}
\def\Sh{{S^{\rm hi}}}

\def\bra{\langle}
\def\ket{\rangle}

\def\khat{\boldsymbol{\hat{k}}}
\def\alm{a_{\ell m}}
\def\Ylm{Y_{\ell m}}

\def\rls{r_{\rm LS}}
\def\tls{t_{\rm LS}}
\def\zls{z_{\rm LS}}
\def\als{a_{\rm LS}}
\def\Hls{H_{\rm LS}}

\def\lcdm{$\Lambda$CDM}
\def\Planck{\textit{Planck}}

\def\camb{{\tt{CAMB}}}

\def\nhat{{\hat{\mathbf{n}}}}

\def\vx{{\boldsymbol{x}}}
\def\vk{{\boldsymbol{k}}}
\def\vd{{\boldsymbol{d}}}

\begin{document}

\title{Testing physical models for dipolar asymmetry: from temperature 
to $k$ space to lensing}

\author{J. P. Zibin} \email{zibin@phas.ubc.ca}
\affiliation{Department of Physics \& Astronomy\\
University of British Columbia, Vancouver, BC, V6T 1Z1  Canada}

\author{D. Contreras} \email{dagocont@phas.ubc.ca}
\affiliation{Department of Physics \& Astronomy\\
University of British Columbia, Vancouver, BC, V6T 1Z1  Canada}

\date{\today}

\begin{abstract}

   One of the most intriguing hints of a departure from the standard 
cosmological model is a large-scale dipolar power asymmetry in the cosmic 
microwave background (CMB).  If not a statistical fluke, its origins must lie 
in the modulation of the position-space fluctuations via a physical mechanism, 
which requires the observation of new modes to confirm or refute.  
We introduce an approach to describe such a modulation in $k$ space and 
calculate its effects on the CMB temperature and lensing.  We fit the 
$k$-space modulation parameters to \Planck\ 2015 temperature data and 
show that CMB lensing will not provide us with enough independent 
information to confirm or refute such a mechanism.  However, our approach 
elucidates some poorly understood aspects of the asymmetry, in particular 
that it is weakly constrained.  Also, it will be particularly useful in 
predicting the effectiveness of polarization in testing a physical modulation.

\end{abstract}

\maketitle


\section{Introduction}
\label{introsec}

   The standard cosmological model, known as $\Lam$ cold dark matter 
(\lcdm), describes the large-scale and early Universe remarkably well, 
with only a handful of parameters (see, e.g., \cite{Planckparams15}).  
Very few hints of departures or tensions with \lcdm\ exist in the present 
cosmological data.  Of these, considerable attention has been paid to 
various so-called ``anomalies'' in measurements of the cosmic microwave 
background (CMB) radiation (see, e.g., 
\cite{WMAPanomalies11,PlanckIandS15,schs15}).  In some cases, the anomalies 
are known to become statistically insignificant when correcting for the 
line-of-sight 
integrated Sachs-Wolfe (ISW) contribution (see, e.g., \cite{emh10,rsd13,rs13}).  
In these cases, due to the weak correlation between the ISW and primary 
anisotropies, the 
anomalies are unlikely to be due to some physical mechanism and hence are 
almost certainly statistical flukes.  However, in all cases the anomalies are 
of only weak to moderate statistical significance, which typically is reduced 
further when correcting for {\em a posteriori} selection effects (also known 
as the ``look elsewhere effect'')~\cite{WMAPanomalies11,PlanckIandS15}.

   One intriguing feature of the CMB temperature ($T$) anisotropies is a 
roughly dipolar power asymmetry~\cite{ehbgl04}.  Measurements with the 
\Planck\ mission~\cite{PlanckIandS15} indicate a roughly $6\%$ amplitude 
of asymmetry up to multipole $\ell \simeq 65$, with a significance (as 
measured by a $p$ value) of roughly $1\%$.  Equivalently, the measured 
amplitude is only about $2$--$2.5$ times the expected level of 
asymmetry due to cosmic variance in statistically isotropic 
skies~\cite{PlanckIandS15}.  The significance of the asymmetry becomes lower 
out to higher $\ell$~\cite{fh13,qn15,awkkf15,PlanckIandS15}, and is reduced to 
of order $10\%$ if we do not consider the scale $\ell \simeq 65$ as predicted 
and correct for {\em a posteriori} effects~\cite{WMAPanomalies11,PlanckIandS15}.

   However, despite its underwhelming statistical significance 
the dipolar power asymmetry remains interesting because of its large-scale 
character.  The asymmetry involves scales that are roughly 
super-Hubble at last scattering, and a number of early-Universe or 
inflationary mechanisms might conceivably affect these scales 
preferentially.  For example, CDM isocurvature fluctuations naturally 
imprint on scales $\ell \lesssim 100$.  However, a particular modulated 
isocurvature model~\cite{ehk09} was recently tested~\cite{Planckinf15} and 
found to not be preferred to \lcdm.

   More generally, it appears to be very difficult to construct a physical 
mechanism  for generating a scale-dependent dipolar modulation (see 
\cite{brst15} for 
a thorough discussion and summary of previous attempts).  This contrasts 
with the relative ease in producing a {\em quadrupolar} modulation (see, 
e.g., \cite{dbmr10,Soda12,nky15}).  The crucial difference is that a 
quadrupolar asymmetry on the sky can be produced via a quadrupolar 
statistical anisotropy 
in $k$ space, associated, e.g., with a homogeneous vector field.  However, 
despite some claims to the contrary~\cite{sh13}, a $k$-space anisotropy 
cannot lead to a dipolar asymmetry on the sky: the reality of the 
fluctuations implies that the $k$-space power spectrum must have even parity 
(see, e.g., \cite{ap10}).  Instead, a dipolar 
asymmetry must be the result of statistical {\em inhomogeneity}, 
perhaps due to modulation with a long-wavelength mode.  Note that this 
distinction holds more generally for any odd compared with any 
even type of asymmetry.  (Parity violation may circumvent this argument; 
see, e.g., \cite{ak16}.)

   It is clear that the important question of whether the observed dipolar 
asymmetry in the CMB temperature fluctuations is due to a statistical fluke 
or to a real, physical modulation of the primordial fluctuations will not 
be resolved through further study of the temperature fluctuations.  This is 
simply because the large-scale $T$ data are already cosmic-variance limited, 
so there will be no significant reduction of noise by remeasuring them.  
What are needed are observations that can probe {\em independent} fluctuation 
modes from those which source temperature.  The most obvious such 
observations are of the CMB polarization.  Although $E$-mode polarization 
is partially correlated with temperature, it is largely sourced by 
independent modes.  Polarization has long been recognized as useful for 
providing independent checks of ``anomalies'' found in the $T$ data (see, 
e.g., \cite{dph08,pgfcdmn10,chss13,gkjr15}).

\begin{figure}
\centerline{\includegraphics[width=\hsize]{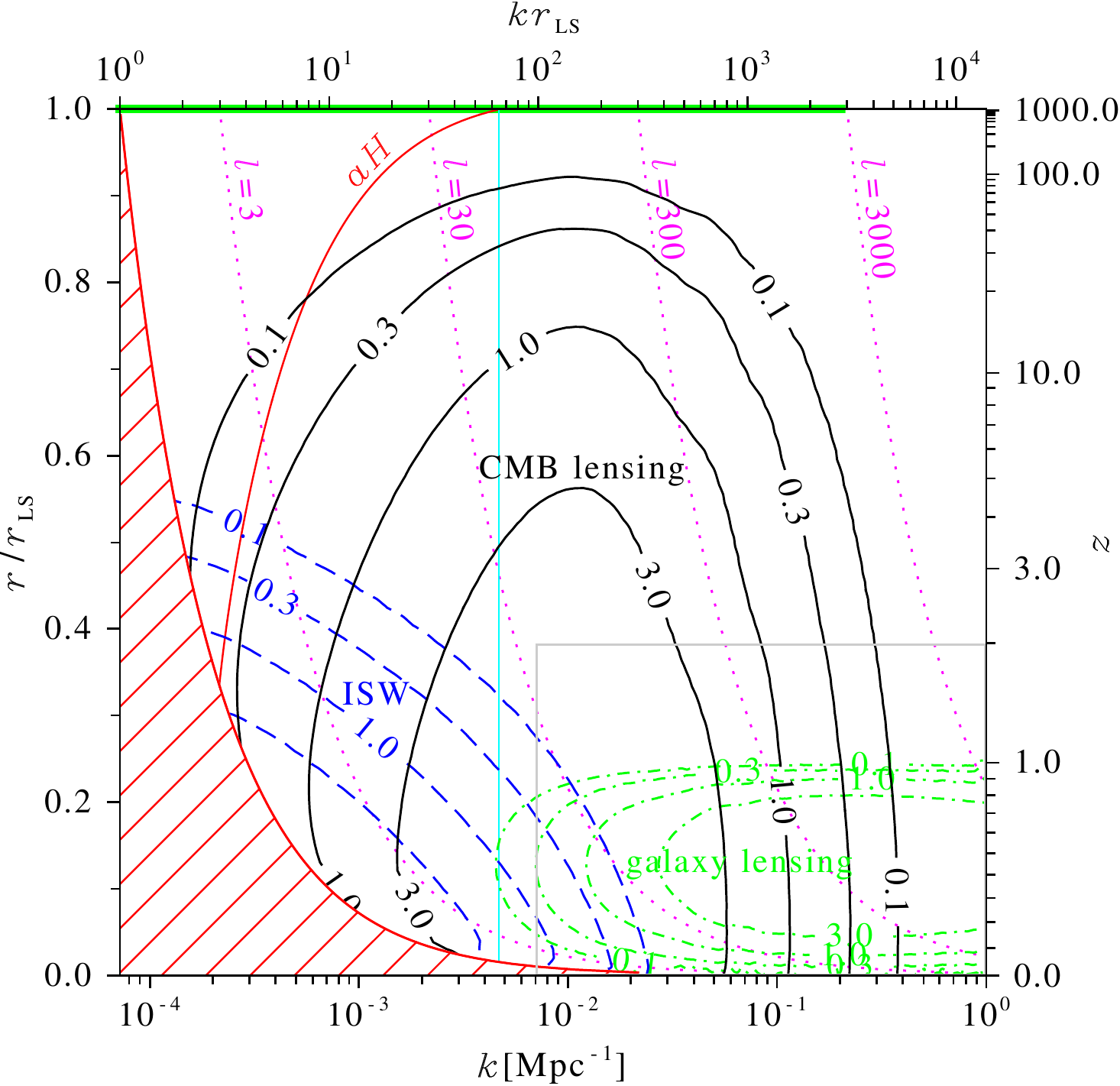}}
\caption{Limber approximation kernels (in arbitrary units) for various 
cosmological observations (contours) out to last scattering ($r = \rls$).  
The grey box indicates very roughly the 
reach of the planned Euclid survey~\cite{euclid09}.  Dotted magenta curves 
correspond to fixed multipole scales, with the red hatched region 
geometrically inaccessible.  The vertical cyan line corresponds approximately 
to the scale $\ell = 65$ in the primary CMB (narrow green box at the top); 
scales roughly to the left of it exhibit dipolar asymmetry in the CMB.  To 
test a modulation model, many more modes are available in principle in our 
observable volume than in the primary CMB source region.  Adapted from 
\cite{zm14}.}
\label{kr_kernel_fig}
\end{figure}

   It is worthwhile considering whether observations other than polarization 
might also be able to address this question.  The essential difficulty is 
that the scales at which the $T$ asymmetry is observed are extremely large.  
To illustrate this, we plot in Fig.~\ref{kr_kernel_fig} the 
Limber approximation kernels 
for various cosmological observations in the $k$-$r$ plane.  (See 
\cite{zm14} for details on the calculations involved.)  The vertical line 
indicates the $k$ scale corresponding approximately to multipole 
$\ell = 65$ in the primary CMB.  We can see that only the ISW effect 
and CMB lensing are currently capable of reaching 
the required large scales.  (Nevertheless, limits on dipolar asymmetry in the 
quasar distribution on much smaller scales were placed in \cite{Hirata09}.)  
However, the ISW effect is mainly sourced at low 
redshifts.  Therefore, for a primordial fluctuation modulation {\em linear} 
in position, the modulation amplitude would be expected to be very small 
for the ISW effect (we will see this explicitly for the case of 
lensing in Sec.~\ref{lenseffectsec}).  In addition the ISW contribution 
mainly appears at the very 
smallest multipoles, so will be heavily affected by cosmic variance.

   Therefore it appears that, after polarization, CMB lensing offers the 
best chance at testing the asymmetry.  However, it should be apparent 
from Fig.~\ref{kr_kernel_fig} that, as with the ISW effect, lensing is sourced 
considerably closer to us than the primary CMB, and hence, for a spatially 
linear modulation, we expect a lower modulation amplitude.  In addition, the 
$k$ scales modulated in the CMB will appear at larger angular scales, i.e.\ we 
expect the asymmetry to appear to lower maximum multipole, in lensing.  
Thus we expect fewer modulated modes for lensing than for temperature.  
For these reasons we expect the significance of detection achievable with 
lensing to be lower than that from temperature.  On the other hand, the 
modes sourcing lensing will be essentially completely uncorrelated with 
the primary CMB temperature, whereas CMB polarization shares significant 
correlation with temperature.

  While most previous studies of the CMB large-scale asymmetry have been 
restricted to $\ell$ or map space, if we observe some amplitude of asymmetry 
out to some multipole scale in temperature we do not expect a CMB lensing 
modulation of the same amplitude and scales, as just explained.  This same 
point will also apply to polarization, due to the different kernels from 
$k$ space to multipole space for these observations.  Therefore, in order 
to obtain predictions for lensing or polarization we must proceed via a 
$k$-space (or position-space) modulation model.

   In this paper we have two main goals.  The first is to present a 
formalism for fitting a $k$-space modulation to CMB $T$ data.  This 
involves first describing a spatially linear modulation in $k$ space, and then 
deriving its effect on the $T$ fluctuations.  We show that this effect 
can be calculated accurately in a very simple way.  We then fit the 
modulation to \Planck\ $T$ data using Bayesian parameter estimation.  Our 
next goal is to determine what the $k$-space model predicts 
for CMB lensing.  To do this we must introduce a formalism for 
calculating the effect of a $k$-space modulation on lensing.

   Our approach will also be applicable to predicting the signal of modulation 
in CMB polarization based on the $T$ observations.  However, besides 
providing such predictions, our rigorous approach to fitting is important 
in its own right.  While grounding the study of the asymmetry firmly in 
$k$ space, we find that temperature data alone are not constraining 
enough to clearly define a $k$-space modulation.  In particular, the 
often-quoted $6\%$ modulation out to $\ell \simeq 65$ does not stand out in 
the data.

  In previous related work, \cite{djkc13} predicted the polarization asymmetry 
given a simplified procedure for fitting to the $T$ data, for modulations of 
various cosmological parameters.  Reference~\cite{naabfw15} considered 
what the $T$ asymmetry predicts for polarization asymmetry via modulated 
primordial spectra, using a similar fitting procedure.  Importantly, they 
found that the polarization predictions are strongly dependent on the 
$k$-space model.  Refs.~\cite{rapj15,kgrkj15} performed more careful fitting, 
but restricted their models.  None of these groups considered lensing.  
Ref.~\cite{fh13} looked for a power asymmetry in the \Planck\ lensing map, 
finding no significant signal in the low-$\ell$ $T$ asymmetry direction.  
Additionally, a recent study~\cite{ms15} claimed that lensing $B$ modes could 
confirm a physical modulation 
at high significance, due to the mode mixing that takes low-$\ell$ lensing 
modes to high-$\ell$ $B$ modes.  However, this paper treated the statistics 
of the lensed $B$ field as Gaussian, whereas it is known that non-Gaussianity 
reduces the total signal-to-noise ratio of the lensing $B$ power spectrum by a 
large factor (see, e.g., \cite{lsc07}).  Also, \cite{ms15} did not consider 
a physical lensing modulation mechanism and simply took the expected 
lensing modulation amplitude to be $7\%$ to $\ell = 70$.

   In this paper we approach this topic in a much more rigorous way.  In 
the first few sections we lay out our modulation formalism.  
Section~\ref{kmodsec} describes our treatment of the $k$-space modulation, 
while Secs.~\ref{Teffectsec} and \ref{lenseffectsec} derive the effects of 
the $k$-space modulation on CMB temperature anisotropies and the lensing 
potential, respectively.  The following sections present our approach to 
fitting the $k$-space modulation to the CMB temperature data 
(Sec.~\ref{fittingsec}), and describe the predicted effect of the modulation 
on the CMB lensing (Sec.~\ref{lenspredsec}).

   Throughout this paper we use the set of \lcdm\ cosmological parameters 
chosen for the Planck Collaboration Full Focal Plane (FFP8) simulations; 
namely, we set Hubble parameter 
$H_0 = 100h\,{\rm km\,s}^{-1}{\rm Mpc}^{-1}$, with $h = 0.6712$, baryon 
density $\Omega_bh^2 = 0.0222$, CDM density $\Omega_ch^2 = 
0.1203$, neutrino density $\Omega_\nu h^2 = 0.00064$, cosmological constant 
density parameter $\Omega_\Lam = 0.6823$, primordial comoving curvature 
perturbation power spectrum amplitude $A_s = 2.09\times10^{-9}$ at pivot 
scale $k_0 = 0.05\,{\rm Mpc}^{-1}$ and tilt $n_s = 0.96$, and optical 
depth to reionization $\tau = 0.065$.  However, we expect our results to 
be only very weakly dependent on these parameters.


\section{Primordial adiabatic $k$-space modulation}
\label{kmodsec}

   Our basic premise is to ask: {\em If} the large-scale CMB temperature 
dipolar asymmetry is due to a real, physical modulation of the primordial 
fluctuations, then what would this predict for CMB lensing (or 
polarization)?  As discussed in the Introduction, a $T$ asymmetry of, say, 
$6\%$ to $\ell \simeq 65$ will not correspond to a lensing (or polarization) 
modulation of 
the same amplitude and angular scales.  To proceed we must specify a form 
for a primordial modulation in position or $k$ space.  This could take the 
form of a modulation of the large-scale adiabatic fluctuations, or 
alternatively a CDM isocurvature or tensor modulation.  The latter two are 
motivated by the fact that they naturally 
give a contribution only on large scales.  Tensor modes, however, are expected 
to produce only tiny gradient-type lensing~\cite{kj97}.  CDM isocurvature 
modes produce considerably less lensing than adiabatic modes, for comparable 
large-scale CMB $T$ contributions.  Therefore, we will restrict our analysis 
here to the modulation of adiabatic modes.  However, when considering the 
predictions for polarization, it will be important to consider these other 
fluctuation types as well~\cite{naabfw15}.

   It is clear that there is no significant {\em scale-independent} dipolar 
asymmetry 
in the CMB temperature fluctuations (see, e.g., \cite{PlanckIandS15}).  Studies 
indicate an asymmetry amplitude of roughly $6\%$ out to multipoles $\ell \simeq 
65$, with decreasing amplitude to larger $\ell$~\cite{PlanckIandS15}.  This 
apparent scale dependence motivates us to treat the primordial adiabatic 
fluctuations as the sum of a large-scale dipole-modulated part and a small-scale 
statistically isotropic part.  The scale dependence of the large-scale part 
will be free, although the total statistically isotropic power will agree 
with $\Lambda$CDM.  In the following we will indicate modulated fields by a 
tilde, while statistically isotropic fields will have no tilde.  We therefore 
write the total primordial (and hence time-independent) comoving curvature 
perturbation, $\wtl\R(\vx)$, as
\beq
\wtl\R(\vx) = \tlRl(\vx) + \Rh(\vx),
\eeq
where the high-$k$ part is statistically isotropic,
\beq
\bra\Rh(\vk)\Rh^*(\vk')\ket = \fr{2\pi^2}{k^3}\PRh(k)\del^3(\vk - \vk').
\eeq
On the other hand, the low-$k$ part is taken to be linearly modulated:
\bea
\tlRl(\vx) &=& \Rl(\vx)\ld(1 + A_\R\fr{r}{\rls}\cos\theta\rd)\label{primmodr}\\
             &=& \Rl(\vx)\ld(1 + A_\R\fr{z}{\rls}\rd),
\label{primmod}
\eea
where $\rls$ is the comoving radius to last scattering, $A_\R$ is a constant, 
the ``modulation amplitude'', and $\theta$ is the angle from the modulation 
direction, which we here define to coincide with the $\hat{z}$ direction.  
$\Rl$ satisfies
\beq
\bra\Rl(\vk)\Rl^*(\vk')\ket = \fr{2\pi^2}{k^3}\PRl(k)\del^3(\vk - \vk').
\eeq
Finally, we take $\Rl$ and $\Rh$ to be uncorrelated,
\beq
\bra\Rl(\vk)\Rh^*(\vk')\ket = 0,
\label{lohiuncor}
\eeq
so that the total statistically isotropic fluctuations, $\R(\vk) \equiv 
\Rl(\vk) + \Rh(\vk)$, must have the usual \lcdm\ power spectrum,
\beq
\bra\R(\vk)\R^*(\vk')\ket = \fr{2\pi^2}{k^3}\PRL(k)\del^3(\vk - \vk'),
\eeq
where
\beq
\PRL(k) = \PRl(k) + \PRh(k).
\label{totalPS}
\eeq
In words, the full-sky ``average'' (or ``equatorial'') power spectrum will 
agree with that of \lcdm\ (at least to lowest order in $A_\R$).  As we explain 
below, treating the fields as two uncorrelated components does not restrict the 
generality of our approach.

   Also, note that we have in mind that $\Rh$ contributes only negligibly to 
the largest scales, so that we expect $A_\R \simeq 0.06$ 
when $\PRl(k)$ extends only to scales corresponding to 
$\ell \simeq 65$, according to the observed $T$ asymmetry.  In this 
study we will take the low-$k$ modulated component to have the spectrum
\beq
\PRl(k) = \half A_s\ld(\fr{k}{k_0}\rd)^{n_s - 1}
          \ld[1 -\tanh\ld(\fr{\ln k - \ln k_c}{\Del\ln k}\rd)\rd].
\label{PRltanh}
\eeq
This spectrum approaches the standard \lcdm\ spectrum for small $k$ and 
approaches zero for large $k$, with cutoff scale $k_c$ and width of cutoff 
$\Del\ln k$.  Recall that the total (isotropic) power spectrum is still 
constrained to have the standard power-law form via \eq(\ref{totalPS}).  
This particular $\tanh$ scale dependence is not intended to model any 
particular mechanism for the modulation of fluctuations.  But it can capture 
some interesting cases.  For large $k_c$, the modulation becomes 
scale-invariant.  By decreasing $k_c$ we can represent a modulation only 
on large scales, e.g.\ scales that are super-Hubble at last scattering, 
which may be related to some early-Universe process.  For $k_c \simeq 
5\times10^{-3}\,{\rm Mpc}^{-1}$ and $\Delta \ln k \rightarrow 0$ in 
particular, we produce a modulation on the commonly quoted angular scales 
of $\ell \lesssim 65$ (keeping in mind that the $k$--$\ell$ kernels imply 
that there is no one-to-one correspondence between $k$ and $\ell$ values).

   The form of the modulation in \eq(\ref{primmod}), i.e.\ that of a spatially 
linearly 
modulated primordial field, is an important assumption here.  We regard it as 
the simplest form that would lead to a dipolar asymmetry.  A linear modulation 
can be considered the lowest-order term in an expansion, for general 
modulations varying slowly on our Hubble scale.  Other choices add 
complexity and require more parameters to specify, e.g., generalizing the 
linear form to quadratic or higher order spatial dependence, or taking the 
fluctuation spectrum to jump like a step function across a ``wall''.  These 
more complicated scenarios could be tested, since they would predict 
asymmetry beyond dipolar, but considering the low signal-to-noise ratio 
of the $T$ asymmetry we restrict this study to the simplest possibility.  
Crucially, the linear modulation means that CMB lensing, which is mainly 
sourced at low redshifts, is expected to be modulated with considerably lower 
amplitude than the observed $T$ amplitude of roughly $6\%$.  This conclusion 
will clearly be strongly dependent on the assumed form of the $k$-space 
modulation.  Also, note that we take the linear modulation to act on the 
{\em primordial} field, $\R$.  This is what would be expected in most proposed 
models where the modulation originates in some very early physics, e.g.\ 
during inflation.  Also, it leads to the linear dependence on {\em comoving} 
distance in \eq(\ref{primmod}).  Conversely, it seems very unlikely that a 
late-time field (e.g.\ the zero-shear gauge fluctuation $\psi_\sig$; see below) 
would be directly modulated.  Such a scenario could involve an 
anisotropic dark energy, which would be subject to strong constraints at the 
background level.  Nevertheless, we will show that, insofar as CMB $T$ and 
lensing are concerned, to a good approximation we can equally consider either 
the early- or late-time fields to be linearly modulated.

   In $k$ space the modulation of \eq(\ref{primmod}) becomes 
\beq
\tlRl(\vk) = \Rl(\vk) + i\fr{A_\R}{\rls}\fr{\pd}{\pd k_z}\Rl(\vk).
\label{Rkmod}
\eeq
This implies that the total $\wtl\R(\vk)$ covariance (to first order in $A_\R$) 
is given by
\bea
\bra\wtl\R(\vk)\wtl\R^*(\vk')\ket
 &=& \fr{2\pi^2}{k^3}\PRL(k)\del^3(\vk - \vk')\nn\\
 &+& 2\pi^2i\fr{A_\R}{\rls}\ld[\fr{\PRl(k)}{k^3} + \fr{\PRl(k')}{k'^3}\rd]\nn\\
 &\times&\del^2(\vk_\perp - \vk'_\perp)\del'(k_z - k'_z),
\label{tlRcovar}
\eea
where $\vk_\perp$ is the projection of $\vk$ orthogonal to $\hat{z}$ and the 
prime on the Dirac delta denotes a derivative with respect to the argument.  
Note importantly 
that, for a Gaussian field $\wtl\R$, \eq(\ref{tlRcovar}) is a complete 
statistical description.  This means that the details of our implementation, 
i.e.\ in terms of the components $\Rl$ and $\Rh$, are irrelevant: in the end we 
obtain a covariance corresponding to a standard isotropic part (the diagonal 
part of \eq(\ref{tlRcovar})), plus a dipole-modulated part with arbitrary 
scale dependence, as determined by $\PRl(k)$ (the off-diagonal, imaginary 
part of \eq(\ref{tlRcovar})).  In particular, our approach does not restrict 
us to some 
early-Universe mechanism which produces two uncorrelated components, $\Rl$ and 
$\Rh$.  The separation into those two components is purely a convenient 
calculational device which will make the analytical work considerably 
simpler, as we will see next.  We remain agnostic as to the physical 
modulation mechanism.  Note that \eq(\ref{tlRcovar}) describes {\em statistically 
inhomogeneous} fluctuations, whereas the effect in $\ell$ or map space will be 
{\em statistical anisotropy.}



\section{Effect on CMB temperature anisotropies}
\label{Teffectsec}

\subsection{Multipole covariance}
\label{ellcovarsec}

   In general, the effect of the modulation, \eq(\ref{primmod}), on the 
CMB anisotropies would be very difficult to calculate (see \cite{kgrkj15} 
for such a general approach).  
However, we will show that, to a very good approximation, the effect will 
be simply to introduce an $\ell$ to $\ell \pm 1$ coupling with spectrum 
determined by $\PRl(k)$, as one might intuitively expect for scales much 
smaller than the length scale of variation of the modulation.

   We begin by demonstrating this on the largest scales, for which we can 
analytically write down the $T$ anisotropies.  Since the observed 
modulation is on large scales, this is a relevant regime.  The large-scale 
approximation used here will begin to break down on scales $\ell \sim 50$, 
although in this case a simple argument will allow us to write down the 
multipole covariance immediately.  Nevertheless, we will provide a detailed 
examination of the small-scale case in the \hyperref[appendix]{Appendix}.

   On the largest scales, it is a good approximation to treat the plasma as 
tightly coupled prior to an instantaneous recombination.  In this 
approximation, the $T$ anisotropies are determined entirely by the 
zero-shear (longitudinal) gauge metric perturbation, $\psi_\sig$, which 
is related to the primordial comoving curvature perturbation, $\R$, via
\beq
\psi_\sig(\vk) = -\fr{3}{5}T(k)\R(\vk),
\label{psiR}
\eeq
where $T(k)$ is the transfer function that captures the effect of radiation 
domination (see, e.g., \cite{zs08}).  Since here we are considering only the 
largest scales, we will ignore the component $\Rh$ in this subsection and 
drop the superscript ``lo'' for brevity.

   Note that in general a linear modulation of $\R$ will not imply a linear 
modulation of $\psi_\sig$, i.e.\ the operations of linear modulation and 
filtering via $T(k)$ will not commute.  An easy way to see this is to 
consider the extreme case of a very narrow filtering around some scale 
$\bar{k}$, $T(k) \simeq \del(k - \bar{k})$.  Then applying $T(k)$ to 
the linearly modulated 
$\R$ will simply give a nearly monospatial-frequency $\psi_\sig$, which 
will not be spatially modulated, as opposed to the case of modulating the 
field filtered with $T(k)$.  Therefore, in general, a linear primordial 
modulation does not lead to a corresponding 
linear modulation of $\psi_\sig$, which is the field that determines the $T$ 
anisotropies.  In practice, this will mean that the calculation of the $T$ 
anisotropies will be very difficult.  On the other hand, for constant 
$T(k)$, the operations of modulation and filtering clearly commute.  So 
as long as $T(k)$ is sufficiently slowly varying, we will be able to 
assume commutativity to good approximation.

   To determine the quantitative effect of the non-commutativity, 
\eq(\ref{Rkmod}) implies
\bea
T(k)\wtl\R(\vk)
   &=& \ld[T(k) - i\fr{A_\R}{\rls}\fr{k_z}{k}T'(k)\rd]\R(\vk)\nn\\
   &+& i\fr{A_\R}{\rls}\fr{\pd}{\pd k_z}\ld[T(k)\R(\vk)\rd].
\label{TRcom}
\eea
Comparing with Eq.~(\ref{psiR}), this tells us that if
\beq
\ld|i\oo{\rls}\fr{k_z}{k}T'(k)\rd| \ll T(k),
\eeq
i.e., if
\beq
\ld|\oo{T(k)}\fr{dT(k)}{dk\rls}\rd| \ll 1,
\label{dTdkrlsT}
\eeq
then the operations of modulation and filtering will essentially commute, so 
that we can write the total $\psi_\sig$ fluctuations to a good approximation 
as linearly modulated according to
\beq
\wtl\psi_\sig(\vx) = \psi_\sig(\vx)\ld(1 + A_\R\fr{r}{\rls}\cos\theta\rd).
\label{psimod}
\eeq
For \lcdm, we find numerically that $T^{-1}(k)dT(k)/d(k\rls) \lesssim 
3\times10^{-3}$ on all scales, so that indeed it will be a very good 
approximation to 
use \eq(\ref{psimod}), which will simplify the calculations tremendously.

   Equation~(\ref{psimod}) makes it very easy to determine the effect of the 
modulation on large-scale anisotropies.  Those anisotropies take the form
\beq
\fr{\wtl{\del T}(\nhat)}{T} = \wtl S(\tls,\rls\nhat),
\label{lgscTmod}
\eeq
for direction $\nhat$ and where $\tls$ is the time of last scattering, and 
the source function $\wtl S(\tls,\rls\nhat)$ is determined fully by 
$\wtl\psi_\sig$ and its first and second derivatives (see, e.g., 
\cite{zs08}).  We have just shown that the linear modulation of $\R$ 
corresponds to very good approximation to the linear modulation of the 
$\psi_\sig$ part of $S(\tls,\rls\nhat)$.  Next we will examine each derivative 
term.  The first spatial derivative takes the form of a radial derivative:
\bea
\oo{\als\Hls}\fr{\pd}{\pd r}\wtl\psi_\sig(\vx)
   &=& \oo{\als\Hls}\fr{\pd\psi_\sig(\vx)}{\pd r}
     \ld(1 + A_\R\fr{r}{\rls}\cos\theta\rd)\nn\\
   &+& \psi_\sig(\vx)A_\R\oo{\als\Hls\rls}\cos\theta.
\eea
The second term on the right-hand side of this expression shows, 
interestingly, that the 
derivative of the modulation gives a term degenerate with the modulation of 
$\psi_\sig$ itself.  However, for \lcdm\ we have $\als\Hls\rls = 66.4$, so 
that this degenerate term can be ignored (for sources near $\rls$) and the 
first derivative of the 
linearly modulated field $\wtl\psi_\sig$ can be well approximated by the 
linear modulation of the derivative of $\psi_\sig$.

   The second spatial derivative takes the form of a Laplacian.  In this case, 
it is trivial that the Laplacian commutes with the modulation in 
\eq(\ref{psimod}), due to the assumed linear nature of the modulation.  The 
same is true for the time derivatives, since the modulation is taken to be 
time independent, as discussed in Sec.~\ref{kmodsec}.  Therefore, the 
temperature anisotropies, \eq(\ref{lgscTmod}), become to a good approximation
\bea
\fr{\wtl{\del T}(\nhat)}{T}
   &=& S(\tls,\rls\nhat)\ld(1 + A_\R\cos\theta\rd)\\
   &=& \fr{\del T(\nhat)}{T}\ld(1 + A_\R\cos\theta\rd).
\label{dTTmod}
\eea
In words, the modulated anisotropies are simply given by the anisotropies 
calculated from the statistically isotropic (``equatorial'') fields, i.e.\ 
$S(\tls,\rls\nhat)$, modulated.

   This leads directly to the simple temperature multipole covariance of the 
form studied in \cite{mszb11}, i.e.\ an $\ell$ to $\ell \pm 1$ coupling.  
Expanding \eq(\ref{dTTmod}) into spherical harmonic multipoles we find
\beq
\wtl\alm = \alm + A_\R\sum_{\ell'm'}a_{\ell'm'}\xi^0_{\ell m\ell'm'}.
\label{modalmlgsc}
\eeq
Here $\xi^0_{\ell m\ell'm'}$ is the polar component of the coupling 
coefficients $\xi^M_{\ell m\ell'm'}$ defined by
\beq
\xi^M_{\ell m\ell'm'} \equiv \sqrt{\fr{4\pi}{3}}
   \int\Ylm^*(\nhat)Y_{\ell'm'}(\nhat)Y_{1M}(\nhat)d\Om_\nhat.
\eeq
Explicitly,
\bea
\hspace{-0.7cm}\xi^0_{\ell m\ell'm'}
   &=& \del_{m'm}\ld(\del_{\ell'\ell - 1}A_{\ell - 1\,m}
    + \del_{\ell'\ell + 1}A_{\ell m}\rd),\\
\hspace{-0.7cm}\xi^{\pm1}_{\ell m\ell'm'}
   &=& \del_{m'm\mp1}\ld(\del_{\ell'\ell - 1}B_{\ell - 1\,\pm m - 1}
    - \del_{\ell'\ell + 1}B_{\ell\,\mp m}\rd),
\eea
where
\bea
A_{\ell m} &=& \sqrt{\fr{(\ell + 1)^2 - m^2}{(2\ell + 1)(2\ell + 3)}},\\
B_{\ell m} &=& \sqrt{\fr{(\ell + m + 1)(\ell + m + 2)}
                        {2(2\ell + 1)(2\ell + 3)}}.
\eea
Equation~(\ref{modalmlgsc}) gives a multipole covariance
\beq
\bra\wtl\alm\wtl a_{\ell'm'}^*\ket = C_\ell\del_{\ell'\ell}\del_{m'm}
   + A_\R\ld(C_\ell + C_{\ell'}\rd)\xi^0_{\ell m\ell'm'}
\eeq
to linear order in $A_\R$, where $C_\ell$ is the power spectrum calculated 
from $\PRl(k)$.  This covariance is a complete statistical description of 
the modulated temperature anisotropies on large scales.

   When the scale of the fluctuations sourcing the anisotropies is much 
smaller than the length scale of variation of the modulation, i.e.\ $\rls$, 
then we would expect the effect of the spatial variation of the modulation 
to be small (see, e.g., \cite{hanson2009}).  In other words, we expect the 
$T$ anisotropies sourced by 
$\PRl(k)$ to be modulated to a good approximation according to 
Eq.~(\ref{dTTmod}).  Nevertheless, it will be worthwhile to be more 
quantitative about this expectation, so we examine small scales in detail 
in the \hyperref[appendix]{Appendix}.

   The simple behaviour for small scales (and the detailed calculations in 
the \hyperref[appendix]{Appendix}) indicate that to very good approximation 
the modulated temperature fluctuations on all scales are given by the 
generalization of Eq.~(\ref{dTTmod}):
\beq
\fr{\wtl{\del T}(\nhat)}{T}
   \simeq \fr{\del T^{\rm lo}(\nhat)}{T}\ld(1 + A_\R\cos\theta\rd)
   + \fr{\del T^{\rm hi}(\nhat)}{T}.
\eeq
\eq(\ref{lohiuncor}) then implies the final result for the multipole 
covariance:
\beq
\bra\wtl\alm\wtl a_{\ell'm'}^*\ket
   = C_\ell^{\Lam{\rm CDM}}\del_{\ell'\ell}\del_{m'm}
   + A_\R(C^{\rm lo}_\ell + C^{\rm lo}_{\ell'})\xi^0_{\ell m\ell'm'}
\label{acovar}
\eeq
to first order in $A_\R$, where $C_\ell^{\Lam{\rm CDM}}$ is the power 
spectrum calculated from $\PRL(k)$ and $C_\ell^{\rm lo}$ is the spectrum 
calculated in the same way but using $\PRl(k)$.

   Notice that the statistical anisotropy in Eq.~(\ref{acovar}) can be 
easily calculated using software such as \camb~\cite{lcl00} with 
the primordial spectrum $\PRl(k)$.  This compares with the approach of 
\cite{kgrkj15} who do not make the approximations we have made and hence must 
calculate some new integrals involving derivatives of internal \camb\ 
variables, which is considerably more work.  Importantly, note that the form of 
Eq.~(\ref{acovar}) is completely general, in that we have the necessary 
standard \lcdm\ form for the statistically isotropic component, and we have a 
dipole-modulated part with a scale dependence that is as free as possible, given 
that it must originate from a $k$-space function (in this case $\PRl(k)$).  
This shows again that our approach of splitting the primordial fluctuations 
into uncorrelated low- and high-$k$ parts, while facilitating the 
calculations, is not restrictive in any way.

   We have ignored the ISW effect in this calculation.  With the linear 
modulation model, the modulation amplitude at the redshifts at which the 
ISW effect is 
sourced is predicted to be considerably smaller (by a factor $r_{\rm ISW}/\rls 
\sim 1/5$) than the roughly $6\%$ for the primary CMB.  Considering also that 
the ISW signal affects mainly the very largest scales, and that smaller 
scales are 
generated at closer distances (recall Fig.~\ref{kr_kernel_fig}), it should be a 
very good approximation to ignore the ISW effect entirely for the asymmetry.  
That is, the 
spectrum $C_\ell^{\rm lo}$ can be calculated without the ISW component.  
Note also that the effect of the 
modulated lensing field on the modulated CMB can also be ignored because it 
is a second-order effect in $A_\R$.

\subsection{Connection to general asymmetry form}

   Using the notation of Ref.~\cite{mszb11}, the general form for the 
multipole moment covariance given a polar ($m = 0$) modulation can be written
\beq
\bra\wtl a_{\ell m}\wtl a_{\ell'm'}^*\ket = C_\ell\del_{\ell'\ell}\del_{m'm}
   + \half\del C_{\ell\ell'}\Del X_0\xi_{\ell m\ell'm'}^0.
\label{acovarAMz}
\eeq
The origin of this notation lies in the assumption that the anisotropy power 
spectrum depends linearly on some parameter, $X$, in which case the modulation 
spectrum, $\del C_{\ell\ell'}$, satisfies
\beq
\del C_{\ell\ell'} = \fr{dC_\ell}{dX} + \fr{dC_{\ell'}}{dX}.
\eeq
This means that we can formally write down the increment in power between 
the modulation equator and the poles as
\beq
\Delta C_\ell = \half\delta C_{\ell\ell}\Delta X_0,
\label{DelCl}
\eeq
where $\Del X_0$ is the change in the parameter $X$ from modulation equator 
to pole.  We will refer to $\del C_{\ell\ell'}$ as the statistically 
anisotropic or modulation power spectrum.

   Comparing Eq.~(\ref{acovarAMz}) to our final result, \eq(\ref{acovar}), 
we can identify
\beq
\Delta X_0 = A_\R
\label{XA}
\eeq
and
\beq
\delta C_{\ell\ell'} = 2\ld(C^{\rm lo}_\ell + C^{\rm lo}_{\ell'}\rd).
\label{dCllp}
\eeq
Equation~(\ref{DelCl}) then allows us to write an effective increment in 
power between the modulation equator and the poles as
\beq
\Delta C_\ell = 2A_\R C^{\rm lo}_\ell.
\label{DelClAR}
\eeq
This is exactly what we would expect, since a fractional modulation of the 
fluctuation amplitude by $A_\R$ should result in a modulation of power by 
$2A_\R$.  This also justifies the approach for calculating the modulated 
$\ell$-space spectra of Ref.~\cite{naabfw15}.

   Note that if there is a significant contribution of $\PRh(k)$ to 
the lowest $\ell$'s, then according to \eq(\ref{DelClAR}) the actual predicted 
asymmetry, $\Delta C_\ell/(C^{\rm lo}_\ell + C^{\rm hi}_\ell)$, will be 
smaller than $2A_\R$.  This is why we said we had in mind that $\PRh(k)$ 
would have a negligible contribution to the largest scales: when this is the 
case our parameter $2A_\R$ will agree well with the actual large-scale 
asymmetry, $\Delta C_\ell/(C^{\rm lo}_\ell + C^{\rm hi}_\ell)$.


\section{Effect on lensing potential}
\label{lenseffectsec}

   In this section we calculate the effect of a linear modulation of the 
primordial fluctuations, $\R$, on the lensing potential.  The (modulated) 
lensing potential is determined by a line of sight integral,
\beq
\wtl\psi^{\rm lens}(\nhat)
= -2\int_0^{\rls} dr\fr{\rls - r}{\rls r}\wtl\psi_\sig(t(r),r\nhat)
\eeq
(see, e.g., \cite{lc06}).  Inserting \eq(\ref{psimod}), which we have shown 
to be an extremely good approximation for the form of the modulated zero-shear 
gauge fluctuations, and using \eq(\ref{psiR}), an expansion in spherical 
harmonics and Bessel functions gives
\begin{widetext}
\beq
\wtl\psi^{\rm lens}(\nhat)
  = \fr{6}{5}\sqrt{\fr{2}{\pi}}\int_0^{\rls}dr\fr{\rls - r}{\rls r}g(t(r))
    \int_0^\infty dkkT(k)
  \sum_{\ell m}\ld[\R^{\rm lo}_{\ell m}(k)
      \ld(1 + A_\R\fr{r}{\rls}\cos\theta\rd) + \R^{\rm hi}_{\ell m}(k)\rd]
  j_\ell(kr)\Ylm(\nhat),
\eeq
where $g(t)$ is the growth suppression factor due to late-time dark energy and
\beq
\R_{\ell m}(k) \equiv i^\ell k\int d\Om_k \R(\vk)Y_{\ell m}^*(\khat).
\eeq
Therefore the lensing potential multipole moments are
\bea
\psi_{\ell m}^{\rm lens}
   &=& \fr{6}{5}\sqrt{\fr{2}{\pi}}\int_0^{\rls}dr\fr{\rls - r}{\rls r}g(t(r))
       \int_0^\infty dkkT(k)\R_{\ell m}(k)j_\ell(kr)\nn\\
   &+& \fr{6}{5}\sqrt{\fr{2}{\pi}}A_\R
       \int_0^{\rls}dr\fr{\rls - r}{\rls r}g(t(r))\fr{r}{\rls}
       \int_0^\infty dkkT(k)\sum_{\ell'm'}\R^{\rm lo}_{\ell'm'}(k)j_{\ell'}(kr)
       \xi_{\ell m\ell'm'}^0.
\label{psilmlens}
\eea
Note the anisotropic part of \eq(\ref{psilmlens}), which contains the $r/\rls$ 
weighting factor.  Finally, we can write the lensing multipole covariance to
$\O(A_\R)$,
\bea
\bra\psi_{\ell m}^{\rm lens}\psi_{\ell'm'}^{\rm{lens}*}\ket
   &=& \fr{144\pi}{25}\int_0^\infty\fr{dk}{k}T^2(k)\PRL(k)
       \ld[\int_0^{\rls}dr\fr{\rls - r}{\rls r}g(t(r))j_\ell(kr)\rd]^2
       \del_{\ell'\ell}\del_{m'm}\nn\\
   &+& \ld[\fr{144\pi}{25}A_\R\int_0^\infty\fr{dk}{k}T^2(k)\PRl(k)\rd.
 \int_0^{\rls}dr\fr{\rls - r}{\rls r}g(t(r))j_\ell(kr)
  \int_0^{\rls}dr'\fr{\rls - r'}{\rls^2}g(t(r'))j_\ell(kr')\nn\\
   &+& \ld.\ph{\fr{1}{1}}(\ell \leftrightarrow \ell')\rd]
       \xi_{\ell m\ell'm'}^0.
\label{covar}
\eea
Using the general definition for the multipole moment covariance given a 
polar ($m = 0$) modulation, \eq(\ref{acovarAMz}), we can identify the 
statistically isotropic part to be
\bea
C_\ell = C_\ell^{\rm lens}
      &=& \fr{144\pi}{25}\int_0^\infty\fr{dk}{k}T^2(k)\PRL(k)
          \ld[\int_0^{\rls}dr\fr{\rls - r}{\rls r}g(t(r))j_\ell(kr)\rd]^2,
\label{Clexact}
\eea
while the statistically anisotropic part is
\bea
\delta C^{\rm lens}_{\ell\ell'}
 &=&         \fr{288\pi}{25}\int_0^\infty\fr{dk}{k}T^2(k)\PRl(k)
    \int_0^{\rls}dr\fr{\rls - r}{\rls r}g(t(r))j_\ell(kr)
   \int_0^{\rls}dr'\fr{\rls - r'}{\rls^2}g(t(r'))j_\ell(kr')
           + (\ell \leftrightarrow \ell'),
\label{Cllpexact}
\eea
\end{widetext}
with $\Delta X_0$ again given by \eq(\ref{XA}).  
The isotropic part, $C_\ell^{\rm lens}$, agrees with the standard 
result~\cite{lc06}, while the anisotropic part, 
$\delta C_{\ell\ell'}^{\rm lens}$, is new.  
It can be easily calculated numerically for \lcdm\ transfer function $T(k)$ 
and growth function, $g(t)$, given a modulation spectrum 
$\PRl(k)$.  Note that unlike the case of the primary CMB anisotropies, for 
lensing the anisotropic part is {\em not} simply the usual lensing spectrum 
calculated with $\PRl(k)$.  The fact that lensing is sourced all along the 
line of sight means that, instead, the last integral in 
\eq(\ref{Cllpexact}) is weighted by a factor of $r/\rls$, which 
reflects the linear nature of the assumed modulation.  As anticipated, this 
reduces the amplitude of the lensing asymmetry relative to that of the 
primary CMB.  The shift to larger angular scales expected for the more 
closely sourced lensing potential is also encoded in \eq(\ref{Cllpexact}).

   We stress that this lensing calculation is considerably simpler than 
that for the temperature fluctuations, due to the simpler relevant transfer 
function and simpler dependence of the lensing potential on the primordial 
fluctuations.  Indeed, the only approximation made here is that of 
\eq(\ref{psimod}), which we have shown to be extremely accurate.

   To complete our description of lensing modulation, we can again formally 
write down the increment in power between the modulation equator and the poles 
as
\beq
\Delta C_\ell^{\rm lens}
   = \half A_\R\delta C^{\rm lens}_{\ell\ell}.
\eeq


\section{Fitting the $k$-space modulation to temperature data}
\label{fittingsec}

\subsection{Formalism}

   Next we describe how we fit the $k$-space modulation spectrum $\PRl(k)$, 
which we have assumed to take the $\tanh$ form of Eq.~(\ref{PRltanh}), to 
CMB temperature data.  The spectrum depends on two free parameters: $k_c$ 
determines which scales are modulated, and $\Delta \ln k$ determines the 
sharpness of the transition from modulated 
to statistically isotropic scales.  We denote these parameters by $p_i = 
\{k_c,\Del\ln k\}$, for brevity.  We begin with the likelihood function for 
the CMB temperature multipoles given the modulation parameters,
\beq
\L(\vd|\Del X_M,p_i) \propto \oo{\sqrt{|C|}}\exp\ld(-\half\vd^\dagger C^{-1}\vd\rd).
\label{eq:likelihood}
\eeq
Here $\vd$ is the vector of multipole moments and the dependence on the 
model parameters $(\Del X_M, p_i)$ is contained in the multipole 
covariance matrix $C$.  Previously we had taken the modulation direction 
to coincide with the $\hat{z}$ direction, but now we must keep the 
direction free and fit for it.  Hence the covariance matrix, 
Eq.~(\ref{acovarAMz}), becomes~\cite{PlanckIandS15}
\bea
\!\!\!\!\!\!\!\!C_{\ell m\ell'm'}
   &\equiv& \bra\wtl a_{\ell m}\wtl a_{\ell'm'}^*\ket\\
   &=& C_\ell\del_{\ell'\ell}\del_{m'm}
    + \half\delta C_{\ell\ell'}\sum_M\Delta X_M\xi_{\ell m\ell'm'}^M.
\label{acovarAM}
\eea
The three model parameters $\Del X_M$ determine the amplitude and direction of 
the modulation (see Eqs.~(\ref{eq:amplitude})--(\ref{eq:phidir}) below), while 
the two modulation parameters $p_i$ determine the scale dependence of the 
modulation via Eq.~(\ref{dCllp}), and so we have in total five parameters 
which describe the 
statistical anisotropy (we hold the main cosmological parameters fixed).

   For fixed $p_i$, we can find the $\Delta X_M$ which maximize the likelihood 
from Eq.~(\ref{eq:likelihood}) to first order in $A_\R$.  Specifically, for 
dipole modulation, we use the estimator from \cite{PlanckIandS15}, which 
generalizes that of \cite{mszb11} (see Ref.~\cite{hanson2009} for related 
optimal estimators):
\begin{align}
  \Delta \tilde{X}_0 &= \frac{6}{f_{10}} \frac{\sum_{\ell m} \delta
  C_{\ell\ell+1} A_{\ell m} S_{\ell m\,\ell + 1\,m}}
  {\sum_{\ell}\delta C_{\ell\ell+1}^2(\ell + 1)F_{\ell}F_{\ell+1}},
  \label{eq:est0}\\
  \Delta \tilde{X}_1 &= \frac{6}{f_{11}} \frac{\sum_{\ell m} \delta
  C_{\ell\ell+1} B_{\ell m} S_{\ell m\,\ell + 1\,m + 1}}
  {\sum_{\ell}\delta C_{\ell\ell+1}^2(\ell + 1)F_{\ell}F_{\ell+1}},
  \label{eq:est1}
\end{align}
and $\Del\tl X_{-1} = -\Del\tl X_1^*$.
Here
\beq
S_{\ell m\ell'm'}
   \equiv T^*_{\ell m}T_{\ell'm'} - \bra T^*_{\ell m}T_{\ell'm'}\ket,
\label{Sdefn}
\eeq
where the $T_{\ell m}$ are $C$-inverse 
filtered temperature multipoles and $F_{\ell}$ is the mean power 
spectrum of the $T_{\ell m}$.  The expectation value in \eq(\ref{Sdefn}) is an 
average over a set of realistic simulations, which provides a mean-field 
correction (described in great detail in 
\cite{Plancklensing14,Plancklensing15,PlanckIandS15}).  The $f_{1M}$ factor 
corrects for normalization errors introduced by masking (its explicit form can 
be seen in \cite{PlanckIandS15}).  The $C$-inverse filter is identical to that 
used in \cite{Plancklensing14,Plancklensing15,PlanckIandS15}, and optimally 
accounts for masking effects.  In practice, we bin the estimator, 
Eqs.~(\ref{eq:est0}) and (\ref{eq:est1}), 
into bins of width $\Delta\ell=1$, which means that the corrections 
to the data described above only need to be calculated once.  This gives exactly 
the same result as if the estimators were computed for each set of $p_i$ from 
scratch; however, it allows us to dramatically speed up the exploration of the 
parameter space (this technique was also employed in \cite{Planckinf15} for 
the same reasons).  
In the following subsection we describe the data 
and corresponding simulations used for obtaining these estimators.  
Given these estimates of the $\Delta X_M$, we can write the best-fit amplitude 
and direction as
\begin{align}
 \tilde{A}_\R &= \sqrt{\Delta\tilde{X}^2_0 + 2|\Delta\tilde{X}_1|^2},
\label{eq:amplitude}\\
 \tilde{\theta} &= \cos^{-1}{\left(\frac{\Delta\tilde{X}_0}{\tilde{A}_\R}\right)},
\label{eq:thetadir}\\
 \tilde{\phi} &= -\tan^{-1}{\left[\frac{\mathrm{Im}{(\Delta\tilde{X}_1)}}
                                       {\mathrm{Re}{(\Delta\tilde{X}_1)}}\right]}.
\label{eq:phidir}
\end{align}

The central limit theorem suggests that the $\Delta X_M$ will be Gaussian 
distributed (this has been verified explicitly with the use of simulations), 
and specifically for statistically isotropic skies they will have mean zero.  
Their variances can be calculated exactly from Eqs.~(\ref{eq:est0}) and 
(\ref{eq:est1}) to be
\beq
\sig_X^2(p_i) \equiv \ld\bra\ld|\Del X_M^2\rd|\rd\ket = \fr{12}
     {\sum_\ell(\ell + 1)\del C_{\ell \ell + 1}^2C_\ell^{-1}C_{\ell + 1}^{-1}}.
\label{gencosvar}
\eeq
The posterior for the $\Delta X_M$ parameters for a fixed $p_i = \bar{p}_i$ is 
then given by
\begin{align}
  P(\Delta X_M, \bar{p}_i | \vd) &= \frac{1}{(2\pi)^{3/2} \sigma_X^3}\nn\\
  &\times\exp{\left[-\frac{\sum_M|\Delta X_M - \Delta \tilde{X}_M|^2}
  {2\sigma_X^2}\right]}.
  \label{eq:delxposterior}
\end{align}

   Using these relations, we can evaluate the log-likelihood function at the 
maximum-likelihood values $\Del\tilde{X}_M$ to be
\beq
\ln\L(\vd|\Del\tilde{X}_M,p_i) = \sum_M\fr{|\Del\tilde{X}_M|^2}{2\sig_X^2},
\label{lnLanis}
\eeq
to first order in $A_\R$ and ignoring terms independent of the statistical 
anisotropy.  This tells us that the expectation of the log-likelihood in 
statistically isotropic skies is independent of $p_i$.  It also says that the 
expected increase of the log-likelihood coming from the introduction of the 
$\Delta X_M$ parameters is 3, as expected.  Note also that this relation 
means that the likelihood will be very simple to evaluate numerically.

   Bayes' theorem allows us to write the posterior for the model parameters as
\beq
P(\Del X_M,p_i|\vd) = \L(\vd|\Del X_M,p_i)P(\Del X_M,p_i),
\eeq
with prior $P(\Del X_M,p_i)$ on the model parameters, up to an overall 
normalization.  We can calculate the posterior marginalized over the $\Del 
X_M$'s, with the result
\beq
P(p_i|\vd) \propto \sig_X^3 \L(\vd|\Del\tilde{X}_M,p_i)P(p_i).
\eeq
Equation~(\ref{lnLanis}) then tells us that a natural choice for the prior on 
the $p_i$ is
\beq
P(p_i) \propto \sig_X^{-3},
\eeq
which yields an expectation of a flat posterior $P(p_i|\vd)$ in statistically 
isotropic skies.  Hence this is the prior we choose.  We also choose a flat 
prior in the $\Del X_M$'s, as is usually done.

   Once the best-fit modulation spectrum parameters $(k_c,\Del\ln k)$ are found, 
it will be a simple matter to evaluate the lensing asymmetry using the method 
laid out in Sec.~\ref{lenseffectsec}, and, in the future, the polarization 
asymmetry as well.

\subsection{Results}

   The results presented here are based on the component-separated temperature 
maps 
provided by the Planck Collaboration~\cite{planckmission}.  Namely, we 
use the {\tt Commander}, {\tt NILC}, {\tt SEVEM}, and {\tt SMICA}\/ 2015 
temperature maps~\cite{planckcompsep} at a {\tt HEALPix}~\cite{Gorski2005} 
resolution of $N_{\rm side} = 2048$ (for brevity we only quote the results for 
{\tt SMICA}; however, we have checked that the other maps do not give 
substantially different results).  We also use the {\tt UT78} mask provided by 
the Planck Collaboration, referred to as the common mask.  We use a set 
of 1000 FFP8 simulations~\footnote{Available on NERSC, at 
\url{http://crd.lbl.gov/departments/computational-science/c3/c3-research/%
cosmic-microwave-background/cmb-data-at-nersc/}}\cite{plancksims}, 
corresponding to each component separation method, in order to make 
mean-field and normalization corrections to the data, as was done in 
\cite{PlanckIandS15}.

Using the relations of Eqs.~(\ref{eq:amplitude})--(\ref{eq:phidir}), we can 
perform a one-to-one linear transformation from the $\Delta X_M$ to Cartesian 
modulation components, $\{\Delta X,\, \Delta Y,\, \Delta Z\}$.  The 
$\{\Delta X,\, \Delta Y,\, \Delta Z\}$ are simply the components of the 
dipole modulation vector in Cartesian Galactic coordinates.  In what 
follows we will present results in this coordinate system for 
convenience.  We scan the model space over 
the following parameter ranges: $\ln(k_c\,[{\rm Mpc}^{-1}]) \in [-7.2, -3.2]$, 
$\Delta \ln k \in [0.01, 0.5]$, 
and $|\Delta X|,\,|\Delta Y|,\,|\Delta Z| \leq 1$.  The lower 
limit on $k_c$ is placed to ensure that we only look for modulation on scales 
that are observable, while the lower limit on $\Delta\ln k$ corresponds 
essentially to an abrupt cutoff in $k$ space.  The upper limits on $k_c$ and 
$\Delta\ln k$ are somewhat arbitrary: in multipole space they correspond 
approximately to limiting the modulation to $\ell < 1000$.  We are primarily 
interested in large-scale modulations, and previous $\ell$-space 
results~\cite{fh13,qn15,awkkf15,PlanckIandS15} indicated no evidence for 
modulation on scales smaller than this limit.  For $A_\R > 1$ the fluctuations 
in \eq(\ref{primmod}) will go to zero somewhere within our last scattering 
surface, and the details of the modulation in this case will depend on the 
specific modulation mechanism.  For the $\tanh$ model we do not approach 
this regime: the limits on the modulation amplitude components turn 
out to be generous.  We explore 
the parameter space using a simple grid approach, which is adequate since 
the parameter space is effectively only two dimensional via 
\eq(\ref{eq:delxposterior}).

\begin{figure*}[ht!]
\begin{center}
\includegraphics[width=\hsize]{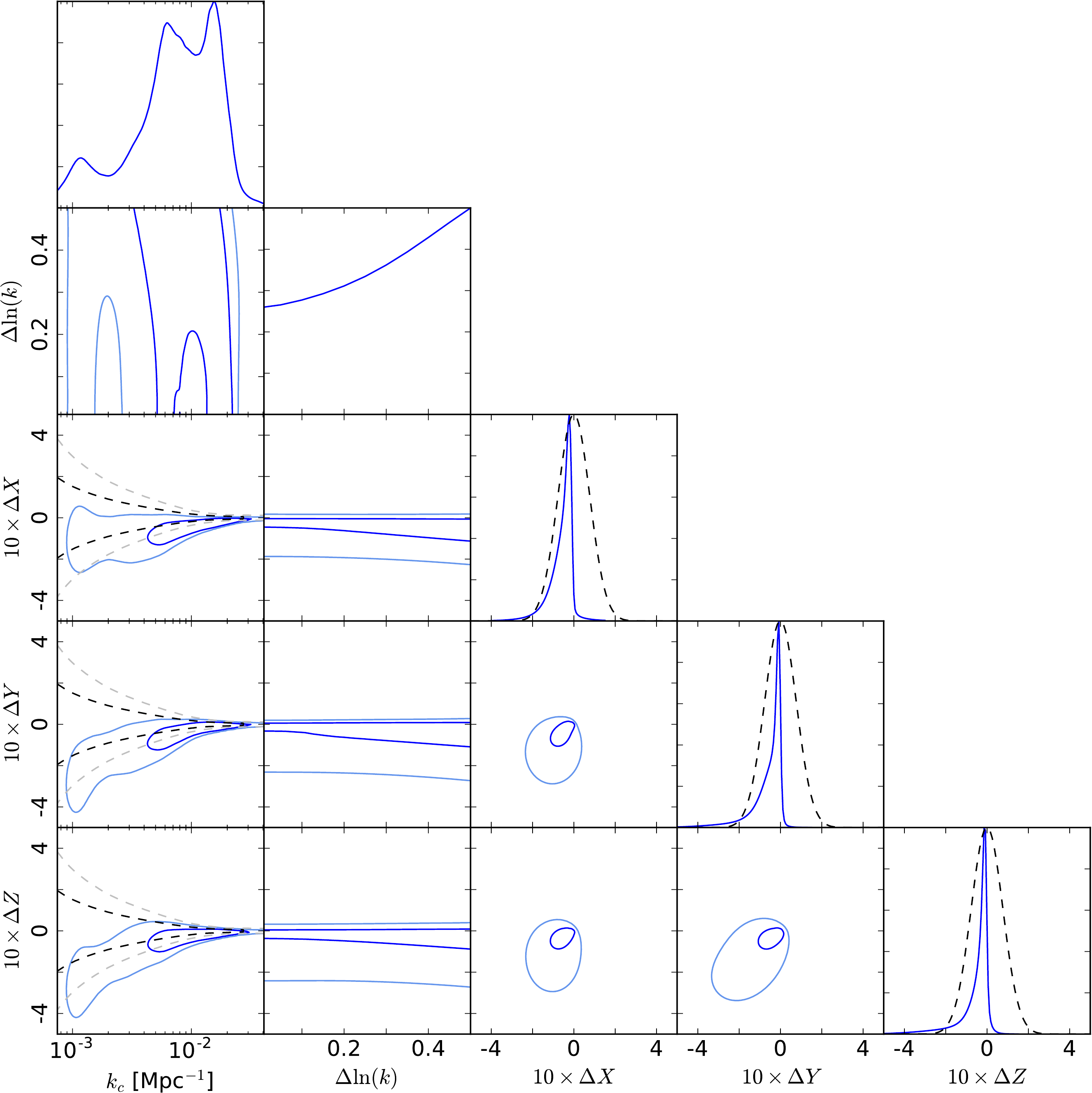}
\end{center}
\caption{Marginalized posteriors for the parameter set $\{p_i, \Delta X, 
\Delta Y, \Delta Z\}$; dark and light blue (solid) contours enclose 68\% and 
95\% of the likelihood, respectively.  The black and grey (dashed) contours 
and curves represent the theoretical distributions of the parameters 
coming solely from cosmic variance in statistically isotropic skies.  The 
values $(k_c, \Delta \ln k) = (5\times10^{-3}\,{\rm Mpc}^{-1}, 0)$ would 
correspond roughly to the often-considered $\ell$-space modulation to 
$\ell \simeq 65$.}
\label{fig:fulltriangleplot}
\end{figure*}

\begin{figure}[h]
\begin{center}
\includegraphics[width=\hsize]{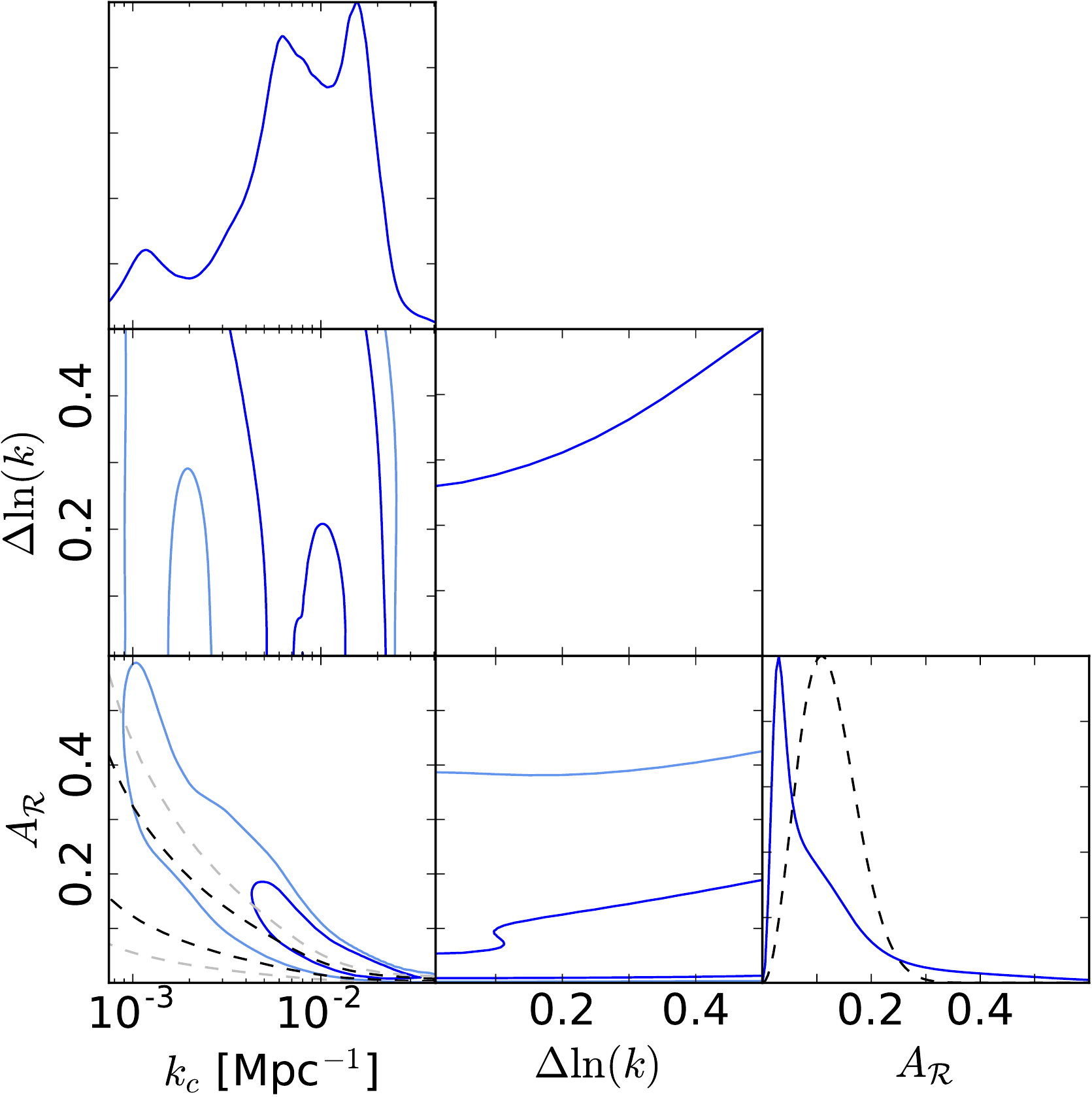}
\end{center}
\caption{Marginalized posteriors for the parameter set $\{p_i, 
A_{\mathcal{R}}\}$; dark and light blue (solid) contours enclose 68\% and 
95\% of the likelihood, respectively.  The black and grey (dashed) contours 
and curve represent the theoretical distributions of the parameters 
coming solely from cosmic variance in statistically isotropic skies.}
\label{fig:triangleplot}
\end{figure}

   Results are summarized as the posterior of the full parameter set 
$\{k_c,\, \Delta \ln k,\, \Delta X,\, \Delta Y,\, \Delta Z\}$ in 
Fig.~\ref{fig:fulltriangleplot}.  We also present results for a condensed 
version of the parameter space, i.e.\ the set $\{k_c,\, \Delta \ln k,\, 
A_{\mathcal{R}}\}$, where the angular variables have been marginalized over, 
in Fig.~\ref{fig:triangleplot}.  We can see from the distributions that no 
parameter is constrained very well.  In particular, $\Delta \ln k$ is completely 
unconstrained by the data, which suggests that there is no well-defined 
transition in the data between modulated and unmodulated scales.  This is not 
surprising, since we have opened up the parameter space in our formalism 
with respect to most previous studies, which considered a sharp cutoff in 
$\ell$ space and only found apparently significant modulation when that 
cutoff was fixed.  
In Table~\ref{tab:bestfitparams} we quote the mean value parameters and 
their uncertainties, which we take as the mean of the marginalized posteriors 
and the area that encloses 68\% of the likelihood.  We also quote the 
maximum-likelihood parameters.  For comparison, when testing for an $\ell$-space 
modulation to $\ell = 65$, Ref.~\cite{PlanckIandS15} found 
$A_\R = 0.062^{+0.026}_{-0.013}$ in the direction $(l, b) = 
(213^\circ,-26^\circ)\pm 28^\circ$.

\begin{table}
\begin{tabular}{cccc}
\hline
\hline
Parameter                   & Mean value                 & Max likelihood\\
\hline
$10^3k_c\,[{\rm Mpc}^{-1}]$ & $7.08^{+12.56}_{-2.34}$    & $7.83$\\
$\Delta \ln k$              & unconstrained              & $0.5$\\
$\Delta X$                  & $-0.060^{+0.054}_{-0.018}$ & $-0.0610$\\
$\Delta Y$                  & $-0.063^{+0.069}_{-0.010}$ & $-0.0414$\\
$\Delta Z$                  & $-0.056^{+0.062}_{-0.004}$ & $-0.0347$\\
$A_{\mathcal{R}}$           & $0.122^{+0.014}_{-0.112}$  & $0.0871$\\
$l\,[^\circ]$               & ${224}^{+43}_{-44}$        & $214$\\
$b\,[^\circ]$               & $-{31}^{+31}_{-16}$        & $-25$\\
\hline
$A_{\mathcal{R}}$           & $0.095^{+0.026}_{-0.080}$  & $\cdots$\\
\hline
\end{tabular}
\caption{Marginalized posterior mean values and their 68\% uncertainties 
for the modulation 
parameters of the model of Eq.~(\ref{PRltanh}), along with their corresponding 
maximum-likelihood values.  The angles $l$ and $b$ are the Galactic longitude 
and latitude, respectively, calculated via Eqs.~(\ref{eq:thetadir}) and 
(\ref{eq:phidir}).  The final row is the combined constraint including an 
ideal CMB lensing experiment assuming a modulation with amplitude 
$A_\R = 0.122$ and the remaining temperature mean values.  The addition of 
lensing does not appreciably help to constrain the model.}
\label{tab:bestfitparams}
\end{table}

\begin{figure}
\includegraphics[width=\columnwidth]{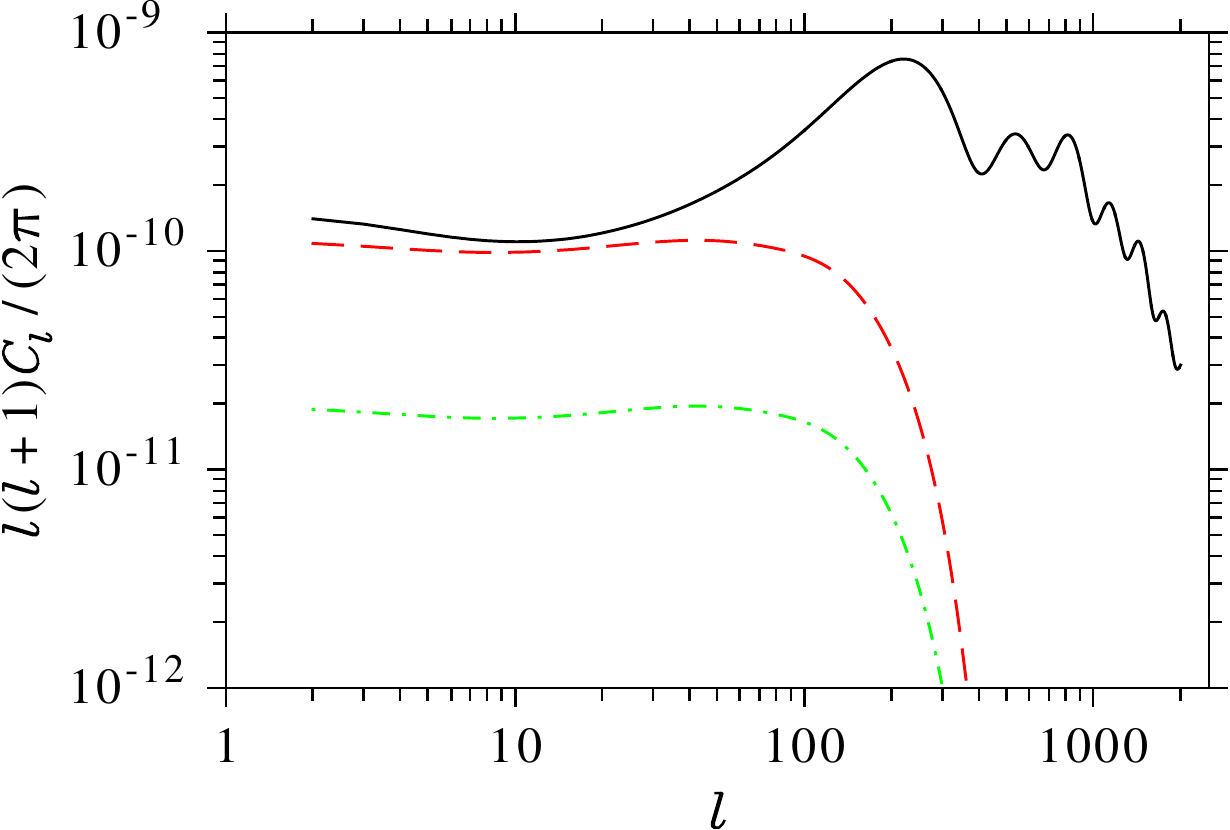}
\caption{Temperature anisotropy isotropic power spectrum, 
$C_\ell^{\Lam{\rm CDM}}$ (solid black curve), anisotropic power spectrum, 
$C_\ell^{\rm lo}$ (dashed red curve), and power spectrum increment from 
equator to pole, $\Delta C_\ell$ (dot-dashed green curve), for the case of 
the maximum-likelihood modulation from Table~\ref{tab:bestfitparams}, 
which fits the observed temperature asymmetry.  The modulation in amplitude 
is at a level of roughly $7\%$ to $\ell \simeq 50$.}
\label{maxLTfig}
\end{figure}

   The temperature anisotropy modulation spectrum $C^{\rm lo}_\ell$ and 
effective power spectrum difference from modulation equator to pole, 
$\Del C_\ell$, for the maximum-likelihood modulation parameters from 
Table~\ref{tab:bestfitparams}, are plotted in Fig.~\ref{maxLTfig}.  These 
were calculated using \camb\ with the corresponding best-fit primordial 
spectrum $\PRl(k)$ and setting the ISW source to zero for redshifts 
$z < 30$.  (Negligible differences were found when the ISW effect was 
included in the 
anisotropic spectrum.)  The modulation in amplitude is at a level of roughly 
$7\%$ to $\ell \simeq 50$.  Importantly, while this agrees crudely with 
the often-quoted level of $6$--$7\%$ to $\ell \simeq 65$, we stress 
that the temperature asymmetry is poorly constrained: the $k_c$ posterior 
in Figs.~\ref{fig:fulltriangleplot} and \ref{fig:triangleplot} has 
significant weight over a large range 
of values, corresponding to $\ell \simeq 50$--$250$.  As the $k_c$-$A_\R$ 
panel in Fig.~\ref{fig:triangleplot} shows, these two parameters are 
anticorrelated, with a larger $k_c$ implying a smaller $A_\R$.  
Furthermore, as the dashed 
contours in that panel show, this anticorrelation follows the trend 
expected from cosmic variance, which arises simply because larger $k_c$ 
implies more modes and hence lower cosmic variance.  
This poorly-defined character of the asymmetry may be surprising, but has 
previously been found in $\ell$ space (see in particular the peaks at 
$\ell \simeq 200$--$300$ in figure 30 of \cite{PlanckIandS15} and figure 15 
of \cite{WMAPanomalies11}, which have similar significance to the peaks at 
$\ell \simeq 65$).

   Finally, note that Fig.~\ref{maxLTfig} shows that an origin to the $T$ 
asymmetry as a modulation of the ISW effect alone (perhaps via an 
anisotropic sound speed for dark energy) is unlikely to produce a good 
fit to the data, 
since the ISW contribution is extremely weak for $\ell \gtrsim 50$.


\section{Predictions for CMB lensing}
\label{lenspredsec}

\subsection{Modulation power spectrum}

   Having used the CMB temperature data to fit the $k$-space modulation 
spectrum, $\PRl(k)$, in the last section, we are now ready to present 
the prediction for the CMB lensing asymmetry.  Once the fitting has been 
done, we know the modulation direction via the $\Del\tl{X}_M$, and so we 
can write the multipole covariance as \eq(\ref{acovarAMz}) with the polar 
direction along the modulation direction and amplitude $\tl A_\R$.

\begin{figure}
\includegraphics[width=\columnwidth]{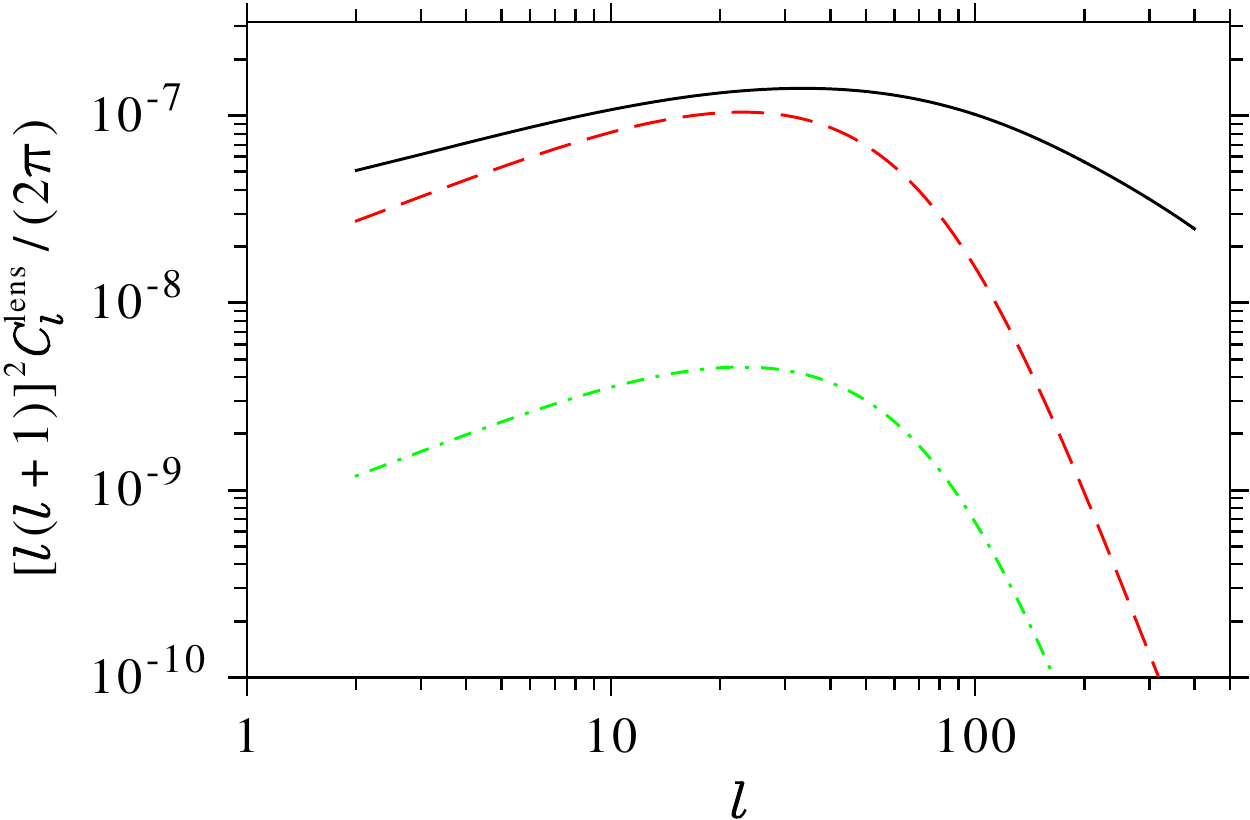}
\caption{Lensing potential isotropic power spectrum, $C_\ell^{\rm lens}$ 
(black curve), predicted anisotropic power spectrum, 
$\del C^{\rm lens}_{\ell\ell}$ (red curve), and predicted power spectrum 
increment from equator to pole, $\Delta C_\ell^{\rm lens}$ (green curve), 
for the case of the maximum-likelihood modulation from 
Table~\ref{tab:bestfitparams}, which fits the observed temperature 
asymmetry.  The modulation in amplitude is at a level of roughly $1.5\%$ 
or less to $\ell \simeq 50$.}
\label{maxLlensfig}
\end{figure}

   Using the maximum-likelihood $k$-space modulation spectrum parameters, 
$k_c$, $\Del\ln k$, and $\Del\tl{X}_M$, from Table~\ref{tab:bestfitparams}, we 
calculated the statistically anisotropic lensing spectrum, $\del C^{\rm 
lens}_{\ell\ell}$, using Eq.~(\ref{Cllpexact}).  The result is plotted in 
\fig\ref{maxLlensfig}.  The lensing spectrum is modulated at a level of about 
$3\%$ and less in power (about $1.5\%$ and less in amplitude), out to scales 
as small as $\ell \simeq 50$.  As predicted in Sec.~\ref{introsec} on 
geometrical grounds, the lensing potential is modulated to a larger minimum 
angular scale and by a smaller amplitude than the corresponding temperature 
best fit presented in Fig.~\ref{maxLTfig}.  This directly leads to a 
low modulation detection significance for lensing, as we will see in the 
next subsection.

   Note that the anisotropic spectrum grows relative to the isotropic 
spectrum at large to intermediate scales.  This can be 
understood with the help of Fig.~\ref{kr_kernel_fig}, where it is apparent 
that larger lensing multipoles are typically sourced at greater distances.  
For our assumed linear modulation form, larger distances, and hence larger 
multipoles, will be modulated with larger amplitude.  Compared with this 
lensing case, the corresponding temperature anisotropy spectrum in 
Fig.~\ref{maxLTfig} exhibits a much more 
similar shape to the isotropic spectrum, up to the cutoff $k_c$ and 
allowing for the lack of the ISW contribution in the 
anisotropic spectrum.  This is simply due to the fact that the primary CMB 
is sourced at essentially a single distance, $\rls$, and hence is modulated 
at a single amplitude for our linear model.

\subsection{Detectability for ideal lensing map}

   In order to ascertain the detectability of the predicted lensing potential 
modulation, we must compare the prediction to the expected uncertainty in the 
measurement.  The expected variance of a CMB lensing measurement of asymmetry 
will be determined by cosmic variance (of the lensing potential modes) and 
lensing reconstruction noise.  It turns out that for a lensing reconstruction 
based on cosmic-variance-limited temperature and polarization anisotropy 
measurements, the reconstruction noise is small compared to the lensing 
potential cosmic variance, at least over the relevant scales~\cite{lppp06} 
\footnote{Note that a lensing reconstruction will be valid even for the 
statistical anisotropies considered here.}.  
Therefore, an ideal lensing measurement can be considered essentially 
cosmic variance limited.  Realistic lensing experiments will have higher 
noise which will necessarily reduce our ability to detect a modulation.  
Hence our conclusions will be conservative.

   We can easily evaluate the cosmic variance of the modulation amplitude 
$\Del X_0$ given a lensing modulation spectrum, 
$\del C_{\ell\ell'}^{\rm lens}$, using Eq.~(\ref{gencosvar}).  
Using \eqs(\ref{XA}), (\ref{Clexact}), and (\ref{Cllpexact}) for 
the case of the maximum-likelihood parameters, we find 
$\sqrt{\ld\bra A_\R^2\rd\ket} = 0.111$.  This means that the 
maximum-likelihood modulation amplitude determined from 
the $T$ anisotropies, $A_\R = 0.0871$, is $0.0871/0.111 = 0.8$ standard 
deviations from zero for a lensing measurement along the known $T$ 
modulation direction.  For the mean value modulation parameters 
from Table~\ref{tab:bestfitparams}, we find an expected measurement of 
$0.9\sigma$.  However, in this case the highly non-Gaussian 
posterior (recall \fig\ref{fig:triangleplot}) means that the mean value 
parameters are biased towards high significance.  In fact, given the 
likelihood from temperature we can determine that the mean detection 
significance for $A_{\mathcal{R}}$ by lensing is $0.7\sigma$ and the 
probability of obtaining a greater than $1\sigma$ detection of 
$A_{\mathcal{R}}$ in lensing is of order 10\%.  
The probability is of order 0.1\% for finding a greater than $1.5\sigma$ 
detection of $A_{\mathcal{R}}$ and decreases quite rapidly for 
higher detection limits.  Therefore, even in this case of an ideal, 
cosmic-variance-limited lensing map, lensing will tell us very little 
about whether the asymmetry is real or not.

   We can illustrate the weakness of CMB lensing for testing a physical 
modulation in another way.  The last row of Table~\ref{tab:bestfitparams} 
lists the result 
of combining the constraint on $A_\R$ from the $T$ anisotropies with the 
expected constraint for lensing, assuming a modulation with parameters 
given by the mean values of Table~\ref{tab:bestfitparams}, which were 
determined by the $T$ likelihood.  
It is apparent that lensing does not improve the constraint 
on $A_\R$ significantly.  If we consider relaxing the condition here that 
the $p_i$ be fixed, we can see that CMB lensing will not be able to 
constrain the modulation model significantly better than CMB $T$ alone.


\section{Discussion}

   In this paper we have presented a rigorous formalism for describing a 
linearly modulated primordial fluctuation field in $k$ space with arbitrary 
scale dependence, and have calculated its effects on CMB temperature 
fluctuations as well as the lensing potential, which probes independent 
modes from the primary CMB.  We performed a Bayesian parameter estimation 
for the $k$-space modulation spectrum, fitting 
to \Planck\ temperature data.  We then predicted the corresponding CMB lensing 
modulation, and found that even an ideal lensing experiment would expect to 
see the modulation at only about $0.7\sigma$.  Hence it appears that CMB 
lensing will never tell us much about whether the observed $T$ modulation is a 
statistical fluke or is due to a real, physical modulation of the primordial 
fluctuations.  Also, this means that the null result for asymmetry in the 
\Planck\ lensing map~\cite{fh13} is completely unsurprising, given that the 
\Planck\ lensing map contains substantially more noise than an ideal map would.

   In principle, correlating CMB lensing with other probes should improve 
the attainable significance of the expected modulation.  However, recall 
from Fig.~\ref{kr_kernel_fig} that current galaxy surveys have weak 
sensitivity at the required extremely large scales.  In addition, such 
surveys reach to relatively low redshifts, and hence we would expect a low 
modulation amplitude, at least for a linear modulation.  Nevertheless, it 
may be worth considering this more carefully, given our result that the upper 
limit for the cutoff is near $k_c \simeq 0.02\,{\rm Mpc}^{-1}$.  The ISW 
contribution reaches to sufficiently large scales, but is sourced so close 
to us that, again, its modulation amplitude is expected to be very small.

   It is important to point out that, although it appears that we cannot 
usefully probe the asymmetry with CMB lensing, it will still be important 
to examine lensing maps for departures from statistical isotropy.  
Lensing probes a large fraction of our observable volume that is 
inaccessible by other means.  Hence it provides a unique opportunity to 
test the simplest models of fluctuations~\cite{zm14}.

   Our results also highlight a seldom-stressed aspect of the temperature 
asymmetry.  We found that no well-defined $k$-space modulation exists, and 
instead that the modulation cutoff scale, $k_c$, is only weakly constrained.  
In particular, there is no reason to single out an approximately $6\%$ 
modulation to $\ell \simeq 65$.  However, this poor 
constraint means that our results should be only weakly sensitive to our 
choice for $\PRl(k)$, i.e.\ to departures from the $\tanh$ form.

   It is clear that polarization will be our best opportunity in the 
near term to test for a physical modulation.  However, even though 
polarization can sample about as many independent modes as temperature, 
Ref.~\cite{naabfw15} finds strong $k$-space model dependence for the 
predictions of polarization.  It will be important to examine this 
with our fitting procedure.  In particular, we will need to generalize 
our approach to incorporate isocurvature and tensor mode modulations.

   In the distant future 21-cm surveys may have the ability to reach to 
large distances and very large scales.  They will have, in principle, 
vastly many more modes within reach via three-dimensional mapping than do 
the two-dimensional CMB or lensing measurements.  Hence they should 
finally resolve the status of the power asymmetry and other anomalies.


\begin{acknowledgments}
   We thank Adam Moss and Ali Narimani for useful discussions.  This research 
was supported by the Canadian Space Agency and the Natural Sciences and 
Engineering Research Council of Canada.  We acknowledge the use of the 
\camb~\cite{lcl00} and {\tt HEALpix}~\cite{Gorski2005} codes.
\end{acknowledgments}

\ph{blank\\}
\ph{blank\\}

   {\em Note added.---}After this work was nearly complete, a related study 
appeared~\cite{hbf15}, which examines the effect of dipole modulation on 
lensing.  That study apparently predicts considerably larger CMB lensing 
modulation amplitudes than we find.  However, it appears that it 
ignores the spatial dependence of the modulation, replacing our 
\eq(\ref{primmodr}) with
\beq
\tlRl(\vx) = \Rl(\vx)\ld(1 + A_\R\cos\theta\rd).
\eeq
Hence they do not see the large reduction in modulation amplitude due to the 
sourcing of lensing at relatively low redshifts.  In addition, \cite{hbf15} 
do not fit a modulation to $T$ data nor do they predict the detectability 
for lensing.

\appendix*
\section{Effect on small-scale $T$ anisotropies}
\label{appendix}

   As we mentioned in Sec.~\ref{ellcovarsec}, when the scale of the 
perturbations sourcing the anisotropies is much smaller than the length 
scale of variation of the modulation, we expect the effect of 
the spatial variation of the modulation to be small.  Nevertheless, it 
will be useful to be more quantitative about this expectation.

   There are three main changes to the temperature anisotropies calculated 
in Sec.~\ref{ellcovarsec} when sources on smaller scales are considered.  
First, the relevant transfer functions become oscillatory in $k$ due to the 
acoustic oscillations.  Next, relaxing the tight-coupling approximation 
means that anisotropic stress must be included.  Finally, relaxing the 
sudden-recombination approximation means that the anisotropies are sourced 
over a range of redshifts, rather than just at $\zls$.  We will consider 
each of these effects in turn.

   To a good approximation, the part of the anisotropy proportional to $\R$ 
(sometimes referred to as the ``monopole'') takes on a term of order (see, 
e.g., \cite{ll09})
\beq
\cos\ld(kr_s\rd)\R(\vk) \equiv T_1(k)\R(\vk),
\eeq
where $r_s$ is the sound horizon.  This term necessarily approaches 
$T(k)\R(\vk)$ in the large-scale limit.  Similarly, the part proportional to 
the radial derivative of $\R$ (the ``dipole'') becomes of order
\beq
\sin\ld(kr_s\rd)\R(\vk) \equiv T_2(k)\R(\vk),
\eeq
which again must approach the large-scale limit $T(k)k/(\als\Hls)\R(\vk)$.  
Recall that we can consider the linear modulation to commute with the 
transfer function filtering if \eq(\ref{dTdkrlsT}) is satisfied.  Here we have
\beq
\ld|\oo{T_1(k)}\fr{dT_1(k)}{dk\rls}\rd| = \fr{r_s}{\rls}\tan(kr_s)
                                      \simeq 0.01\tan(kr_s)
\label{T1com}
\eeq
and
\beq
\ld|\oo{T_2(k)}\fr{dT_2(k)}{dk\rls}\rd| \simeq 0.01\cot(kr_s).
\label{T2com}
\eeq
Therefore for most $k$ scales, the condition for commutativity is met.  
For the dipole term, the $\cot$ dependence may suggest a problem as 
$k \rightarrow 0$.  However, we showed explicitly in Sec.~\ref{ellcovarsec} 
that the dipole term does, in fact, commute to a good approximation with 
modulation on large scales.  Similarly, the periodic divergences in 
$\tan(kr_s)$ and $\cot(kr_s)$ at larger $k$ values may suggest that 
commutativity breaks down at these scales.  To examine the effect of these 
divergences, consider the covariance of $T(k)\wtl\R(\vk)$ calculated 
using Eq.~(\ref{TRcom}) for $T(k) = \cos(kr_s)$.  In 
addition to the expected statistically isotropic term proportional to 
$\cos^2(kr_s)\PR(k)$, we find an extra isotropic term proportional to 
$\cos(kr_s)\sin(kr_s)\PR(k)r_s/\rls$.  Very close to the zeros of 
$\cos(kr_s)$ this extra term will dominate.  However, its {\em absolute} 
contribution is weighted by the small factor $r_s/\rls$.  The relatively 
broad kernel that takes us from $k$ to $\ell$ space will mean that the 
extra term will alter the acoustic peak structure only by a small amount, 
in proportion to the factor $r_s/\rls$.  This tells us that the 
modulation commutes to good approximation with the acoustic oscillation 
processing.

   The next small-scale effect is the presence of anisotropic stress, i.e.\ 
the quadrupole Boltzmann terms.  These terms are suppressed by factors 
$k/|\dot\tau| \sim 10^{-3}k\rls$, where $\tau$ is the optical depth (see, 
e.g., \cite{hw97}).  The anisotropic stress is sourced within distances of 
the order the mean free path from the observed point on the last scattering 
surface, which is much smaller than $\rls$, and is determined by gradients 
of the primordial field.  Hence, as we showed for the case of the derivative 
terms in Sec.~\ref{ellcovarsec}, for these contributions modulation will 
commute to a very good approximation with filtering.  Importantly, as 
polarization is sourced entirely by anisotropic stress, this will mean that 
we will be able to describe in a similar way the effect of modulation on 
polarization.

   The final small-scale effect is the sourcing over a range of redshifts, 
weighted by the visibility function.  Including also the high-$k$ part of the 
fluctuations, the anisotropy of \eq(\ref{lgscTmod}) becomes in this case the 
line-of-sight integral
\beq
\fr{\wtl{\del T}(\nhat)}{T}
   = \int_0^\infty dr\ld[\wtl\Sl(t(r),r\nhat) + \Sh(t(r),r\nhat)\rd].
\eeq
The previous arguments tell us that, to a good approximation, we can write
\beq
\wtl\Sl(t(r),r\nhat)
   \simeq \Sl(t(r),r\nhat)\ld(1 + A_\R\fr{r}{\rls}\cos\theta\rd).
\eeq
Therefore, writing $r = \rls + \del r$, the anisotropy becomes
\bea
\fr{\wtl{\del T}(\nhat)}{T} \simeq \int_0^\infty&&\hspace{-0.4cm}dr
     \ld[\Sl(t(r),r\nhat)\ld(1 + A_\R\cos\theta\rd)\rd.\nn\\
  &+& \ld.\Sh(t(r),r\nhat)\rd] + \O(\del r/\rls).
\eea
Since the primary anisotropies are sourced over a range of distances 
$\del r/\rls \sim 10^{-3}$, we have to very good approximation
\beq
\fr{\wtl{\del T}(\nhat)}{T}
   \simeq \fr{\del T^{\rm lo}(\nhat)}{T}\ld(1 + A_\R\cos\theta\rd)
   + \fr{\del T^{\rm hi}(\nhat)}{T}.
\eeq
Using \eq(\ref{lohiuncor}), this leads immediately, as in 
Sec.~\ref{ellcovarsec}, to the final result for the multipole covariance:
\beq
\bra\wtl\alm\wtl a_{\ell'm'}^*\ket
   = C_\ell^{\Lam{\rm CDM}}\del_{\ell'\ell}\del_{m'm}
   + A_\R(C^{\rm lo}_\ell + C^{\rm lo}_{\ell'})\xi^0_{\ell m\ell'm'}
\eeq
to first order in $A_\R$, where $C_\ell^{\Lam{\rm CDM}}$ is the power 
spectrum calculated from $\PRL(k)$ and $C_\ell^{\rm lo}$ is the spectrum 
calculated in the same way but using $\PRl(k)$.

\bibliography{klensmod}

\end{document}